TITLE

Ultrashort 30-*fs* laser photoablation for high-precision and damage-free diamond machining

AUTHORS

Maksym Rybachuk [a, b]*, Bakhtiar Ali [a, b], Igor V. Litvinyuk [b, c]

AFFILIATIONS

[a]      School of Engineering and Built Environment, Griffith University, 170 Kessels Rd., Nathan QLD 4111, AUSTRALIA

[b]      Queensland Quantum and Advanced Technologies Research Institute, Griffith University, 170 Kessels Road, Nathan, QLD, 4111, AUSTRALIA

[c]      School of Environment and Science, Griffith University, Nathan QLD 4111, AUSTRALIA

CORRESPONDENCE

* Corresponding author:

m.rybachuk@griffith.edu.au

https://orcid.org/0000-0002-5313-9204

ABSTRACT

A 30-*fs*, 800 nm, 1 kHz femtosecond was used to photoablate diamond across radiant energy doses of 1 - 500 kJ/cm², with fluences of 10 - 50 J/cm² and, pulse counts from 100 to 10,000. The objective was to maximise material removal while minimising surface roughness ($R_a$) by operating above the photoablation threshold. Results demonstrate that 30-*fs* laser photoablation achieves $R_a$ <0.1 $\mu$m, meeting both high- and ultra-high-precision machining standards, while maintaining surface integrity and preventing heat-affected zone (HAZ) damage. At 1 kJ/cm² (10 J/cm² fluence, 100 pulses), an $R_a$ of 0.09 $\mu$m was achieved, satisfying ultra-high precision criteria ($R_a$ <0.1 $\mu$m). Additionally, doses below 10 kJ/cm² consistently met high-precision machining requirements ($R_a$ <0.2 $\mu$m). Photoablation efficiency peaked below 50 kJ/cm², after which material removal diminished, indicating non-linear process limitations. The $sp^3$ diamond phase remained intact, as confirmed by the unchanged $T_{2g}$ Raman mode at 1332 cm$^{-1}$, with no detectable Raman $G$ or $D$ modes, confirming the absence of $sp^2$-related graphitization, structural disorder, of nitrogen vacancy (NV) centre annealing. These findings establish 30-*fs* laser processing as a high-precision, damage-free approach for diamond machining, with promising applications in NV centre-containing quantum materials and advanced tooling.

# 1    INTRODUCTION

The demand for miniaturized opto-electronic and micro-mechanical systems has necessitated the need for both high-precision and ultra-high-precision finished components made of diamond [1] [2], including those fabricated using diamond as a single-point cutting tool in diamond turning lathes [3]. The latter machine photonic components, optical mirrors, x-ray optics, scanner mirrors, infra-red lenses and many other critical opto-electronic elements [4]. Fabrication of these components must satisfy certain average arithmetic surface roughness ($R_a$) values set by the ISO 1302 [5] and ASME B46.1 [6] standards relative to specific ISO 286-1 [7] tolerances, requiring an $R_a$ of ≤0.1 $\mu m$ for ultra-high-precision, and an $R_a$ of ≤0.2 $\mu m$ for high-precision machined components. Additionally, for high-precision machining, including difficult-to-machine advanced ceramics and carbides, a material removal rate (MRR) exceeding 0.001 mm³/s is generally considered as acceptable [8], whereas for ultra-high-precision machining an MRR of 0.0001 mm³/s is acceptable [9].

Conventional subtractive methods, including cleaving, sawing, and mechanical polishing, are often unsuitable to achieve such high surface finishes, particularly for complex geometries. Laser-based micro-machining offers a viable alternative, however traditional nano- (*ns-*) and pico- (*ps-*) second pulsed laser photoablation deliver excessive radiant energy within a single pulse, ionizing both electrons and atoms through thermal processes, resulting in significant heating within the focal volume due to phonon-driven lattice diffusion, which causes material melting and unavoidable damage through heat diffusion and formation of heat-affected zones (HAZs) [10] in processed workpieces. Additionally, the *ns-* and *ps-* irradiation effects are detrimental to *sp³*-to-*sp²* phase conversion in diamond as these reduce crystalline quality of photoablated surfaces by forming hard amorphous olefinic *sp²* and amorphized *sp³* fractions in the HAZs [11].

Ultra-short femtosecond (*fs*) laser pulses, lasting only a few 10s of *fs*, provide a solution by enabling a HAZ-free ablation as *fs-*pulses deliver energy in a timeframe at least an order of magnitude (×10¹ s) shorter than the electron-lattice phonon diffusion time [12]. With *fs* laser peak powers reaching several MWs at total energy per pulse remaining low, *fs-*laser field, unlike *ns-* and *ps-*lasers, interacts with solid state electrons only, inducing multi-photon [13] and tunnelling ionization via distortion of Coulomb potential [14], and thus making the photoirradiation entirely non-thermal, offering precise photoablation with minimal heat damage [15, 16]. Unlike *ns-* or *ps-*lasers, *fs-*pulsed lasers may effectively prevent incomplete *sp³*-to-*sp²* phase conversion and avoid the formation of olefinic *sp²* and amorphized *sp³* fractions in diamond. However, even *fs-*pulses as short as 100 *fs* induce electron (*e⁻*) thermalization, generate hot ions and phonons, leading to thermal relaxation within the lattice [17, 18] that (may) compromise machining precision. Recent studies[16, 19] suggest that sub-50 $fs$ pulses could eliminate these limitations, yet their potential for diamond micromachining remains largely unexplored.

To optimize $fs$-laser processing, it is crucial to establish the appropriate energy dose (*i.e.,* the total amount of radiant energy, kJ/cm²), which is the product of the fluence (energy per unit area, J/cm²) and the number of pulses ($n$) per irradiation site, delivered to a target area [20]. The fluence must exceed materials ablation threshold, which for diamond's is reported at ~3 J/cm² for 100 $fs$, 800 nm irradiation [21], while balancing MRR and $R_a$. Achieving an optimal above-threshold dose is critical to improving precision, efficiency, and surface integrity in laser machining of difficult-to-machine wide band gap materials, such as diamond.

This study investigates the feasibility of 30-$fs$ laser photoablation for high-precision and ultra-high-precision diamond machining by addressing three (3) fundamental questions:

1.      How does 30-$fs$ pulsed laser irradiation affect the surface morphology and $R_a$ of diamond when operating well above its ablation threshold?

2.      What energy dose optimizes $R_a$, MRR, and bulk volume removal for high-precision and ultra-high-precision diamond machining?

3.      Does 30-$fs$ laser processing introduce measurable undesirable structural defects, $sp^3$ disorder, or HAZs in the processed diamond?

By addressing these questions, our work attempts to provide critical insights into the feasibility of application of ultrashort 30-$fs$ laser process as a damage-free, high-precision alternative to conventional longer pulse $ns$- or $ps$- laser irradiation processes.

## 2      EXPERIMENTAL DETAILS

### 2.1      Materials

An $n$-type <100> CVD type-Ib diamond samples (3×3×1 mm; Chenguang Machinery Co. Ltd., Hunan, China) with a nitrogen, N, content of 200 ppm and, an $R_a$ of 40 nm, measured on planar surfaces prior to all experiments, were chosen for the study. The nominal $R_a$ of CVD diamond sample surface was *ca*. 30 nm as reported by the manufacturer for all samples tested.

### 2.2      Laser photoablation set up and design of experiments

A custom-made Ti:sapphire laser system (FEMTOPOWER™ compact™ PRO HE, FEMTOLASERS Produktions GmbH, Austria) delivering a linearly polarized Gaussian shaped ultrashort 30 *fs* laser pulse at a pulse repetition rate, *f*, of 1 kHz, and a central wavelength of 800 nm, with each pulse carrying 0.8 mJ, was used in the study. A CMOS camera beam profiler with 2448 × 2048 px resolution (BC207VIS(M), Thorlabs Inc., USA) and a protected aluminium 90° off-axis parabolic (OAP) mirror (MPD149-P01, Thorlabs Inc., USA) were used to focus the beam onto the sample surface, with a focal spot diameter, $D_f$, of ~15 $\mu$m at 1/e of the maximum intensity. Schematic layout of the system comprised of a 30 *fs* 800 nm laser source, a rotating ½- wave plate, Glan-Taylor prism polarizers, a beam splitter (for beam profiling), a neutral density (ND) filter, an aperture and a sample translational stage, is shown in Fig. 1.

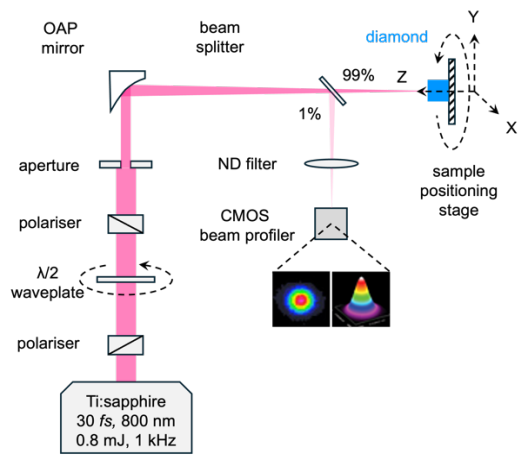

Figure 1. Schematic layout of a 30-*fs* laser photoablation system for CVD diamond processing.

The fluence for all irradiation experiments was set above the highest *fs* reported ablation threshold of diamond (~3.0 J/cm²), ranging from 10 to 50 J/cm² at 10 J/cm² increments. Pulse energy was adjusted through a combination of two mounted wire grid polarizers (WP25M-UB, Thorlabs Inc., USA) and a half-waveplate, to achieve the desired fluence. A quarter waveplate was used to convert the beam polarization from linear to circular, aiming to achieve improved $R_a$ with the laser beam, controlled by a motorized precision translation stage, scanning across the diamond surface. The design of experiments (DoE) included the set pulses counts, *n*, varied from 100 to 10,000, roughly in a non-linear (logarithmic) fashion, 100 → 500 → 1000 → 5000 → 10000 to capture possible non-linear MRR and $R_a$ trends across the *n* range. The range of *n*, in the DoE was chosen to capture non-uniform (*e.g.,* sub-linear or super-linear) changes across the energy doses from 1 to 500 kJ/cm$^2$ without needing to collect data at every possible linear interval as diamond's $sp^3$-to-$sp^2$ phase conversion is known to be gradual. Fewer data points at higher $n$ were selected to avoid experimental oversampling as at 5,000 and 10,000 $n$ the phase changes are expected to stabilize, whereas at 100 and 500 $n$ capture early-stage trends, with the results expected to vary more significantly with smaller increases. At constant pulse repetition rate, $f$, and focal spot diameter, $D_f$, lower $n$ corresponds to lower radiant energy density per irradiation site, and *vis versa*.

## 2.3 Sample post-processing following laser photoablation

After laser irradiation, samples were rinsed with 70% nitric acid (HNO$_3$, ACS reagent, Sigma-Aldrich, Merck Group, USA) and sonicated in an ultrasonic cleaner (Skymen JP-100plus, Shenzhen, China) at 25 kHz, and 800 W for 6 hrs to remove $sp^2$ graphitized inclusions. The samples were naturally air-dried.

## 2.4 Surface topography measurements and approximations

High-resolution three-dimensional (3D) surface topography data were extracted from the shallow linear photoirradiated regions (aka. tracks) in diamond using the optical profiler (Zeta™ 300, KLA Corporation, USA) equipped with surface interferometry probes, including the phase scanning interferometry (PSI) and vertical scanning interferometry (VSI), combined on the piezo stage and interferometric objective lens. The PSI measured step heights from a few Å to 100s of $n$ms, while the VSI covered ranges from 100s of $n$ms to 100s of $\mu$ms recording $R_a$ formed on channel walls and, at the bottom of the machined channel, and features such as ridge, edge, and surface pile ups, burrs or cracks withing the photoablated regions. The 3D interferometric depth, $d$, and width, $w$, values were used to estimate MRR, adjusted for sample translational velocity in accordance with the earlier reports by Ali *et al.* [20]. Each 3D interferometric measurement was repeated three (3) times, with the average value for $d$ and $w$, in $\mu$m, reported as the final output, minimizing measurement errors. The irradiated volume ($V$, $\mu$m$^3$) removed was estimated using a cone approximation with the cone volume scaled by a factor of 1.19 (19% larger) to approximate the Gaussian function. This adjustment accounts for the Gaussian beam profile over a finite domain (*i.e.,* within $\pm \sigma$) using a base radius (1/2 $D_f$) and height (used as $d$) of the Gaussian's beam. The MRR was estimated by considering the effective irradiated volume, in $\mu$m$^3$, removed by laser irradiation over the exposure time, $s$ [20].

## 2.5    Raman spectroscopy measurements

Raman spectra across the $800 - 2200$ cm$^{-1}$ range were obtained using an unpolarized 532 nm Ar$^+$ ion laser confocal micro-Raman microscope (Renishaw inVia™, UK) equipped with WIRE® 3.4 software and a ×50 objective, producing a focal spot ~1.5 $\mu$m. The measurements were taken at 293 K at 1 cm$^{-1}$ resolution, over a 60 $s$ acquisition time, using $\leq$ 1 mW laser power to minimize secondary thermal damage to the samples [22]. Unmodified Raman spectra showing a linear photoluminescence (PL) background were processed using MagicPlot Pro® 3.0.1 (2022 Magicplot Systems, LLC.).

## 3    RESULTS AND DISCUSSION

### 3.1    Diamond surface profile

Figure 2(a) presents a 3D micrograph of a photoablated linear region, while Figure 2(b) shows its cross-sectional profile, for a CVD diamond sample irradiated with 30-$fs$ laser pulses at a 2 kJ/cm² energy dose (20 J/cm² fluence, 100 pulses). The 3D micrograph shows a triangulated cross-section owing to the inverted Gaussian distribution of laser peak intensity, with the peak maxima aligned to the centre of the channel. The sample was irradiated at a fluence ~10× higher than the $fs$-laser ablation threshold of diamond (3.0 J/cm²), however, no HAZs, skirting, or pile-up features were observed in the photoablated profile, indicating high machined surface quality. This is attributed to the non-thermal nature of sub-50 $fs$ photoablation [17, 23].

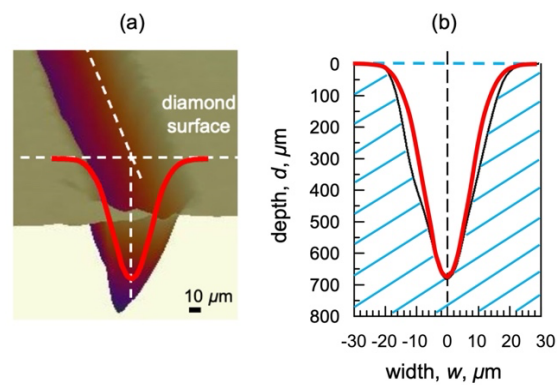

Figure 2: (a) a 3D micrograph, (b) cross-sectional profile of a diamond photoablated region irradiated with 30 $fs$-laser pulses at a 2 kJ/cm$^2$ energy dose (20 J/cm$^2$ fluence, 100 pulses). An inverted Gaussian function (red) is superimposed on the profile.

Laser-ablated channels displayed slight asymmetry and uneven walls, influenced by several factors, including (i) optically induced aberrations from the distorted Gaussian beam profile as it is shaped through the optical alignment system and parabolic mirror (see Section 2.2), (ii) variations in diamond absorption and potential misalignment of the sample surface normal to the beam, leading to non-uniform energy absorption, and (iii) translational stage motion, which may have affected beam projection, contributing to focal asymmetry. Achieving more uniform photoablation requires precise beam delivery and sample positioning to minimise asymmetry and improve uniformity of laser-ablated regions.

## 3.2    Width and depth profile of photoablated tracks

The variation of photoablated track width, $w$, and depth, $d$, in $\mu$m, are shown in Figure 3(a) and Figure 3(b), respectively, as a function of applied laser energy dose, kJ/cm$^2$. A distinct transition threshold was observed at a relatively low irradiation dose (~50 kJ/cm$^2$), beyond which both $w$ and $d$ increased almost exponentially. This threshold energy can be reached through various fluence and $n$ combinations (see Supplementary Figure S1).

Figure 3 shows that $d$ responds non-linearly to increasing $fs$-laser dose, unlike $w$. In the low-dose regime (<50 kJ/cm²) where multi-photon ionization governs sublimation, $d$ is primarily shaped by the Gaussian beam profile, where $w$ increases as more material sublimes from the periphery of the beam. As the dose increases, $d$ responds non-linearly due to plasma shielding and distorted energy dissipation in the ablated volume [24]. In the high-dose regime (>50 kJ/cm²), excess energy spreads laterally, rather than penetrating deeper, limiting further depths improvement. Consequently, 30 $fs$ laser doses in excess of 50 kJ/cm$^2$ provide minimal gains in ablation depth due to energy redistribution.

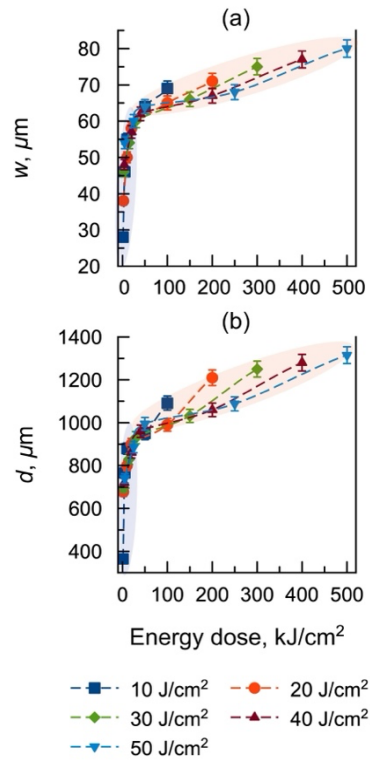

Figure 3: The variation of photoablated track (a) width, *w*, and (b) depth, *d*, in $\mu$m in diamond as a function of applied 30-*fs* laser energy dose. Low-dose (<50 kJ/cm²; blue) and, high-dose (>50 kJ/cm²; orange) ablation regime are indicated with overlays.

Detailed changes in 3D ablation profile are shown in Figure 4, as a function of applied fluences from 10 to 50 kJ/cm², and *n* across 100 - 10,000 region shown in Figure 4(a) and, with *n* limited to 100-pulse region shown in Figure 4(b). The smallest *w* and *d* values are observed at the lowest dose of 1 kJ/cm² (10 J/cm² fluence, 100 *n*). However, *d* increased sharply from ~350 $\mu$m to ~700 $\mu$m, as the fluence doubled from 10 J/cm² to 20 J/cm² at constant 100 *n*, corresponding to a dose increase from 1 kJ/cm² to 2 kJ/cm².

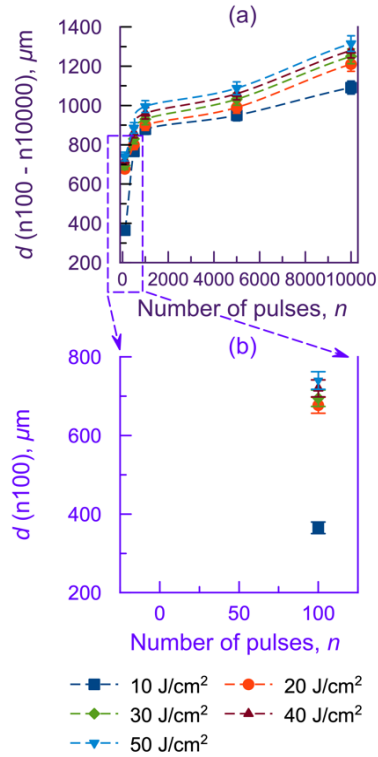

Figure 4: The variation of photoablated track depth, *d*, in *µ*m in diamond as a function of applied 30-*fs* laser fluences from 10 to 50 kJ/cm², and pulse numbers per irradiation site across the (a) 100 to 10,000 pulses, and (b) within the 100-pulse region.

A rapid two-fold increase in the laser-machined features between 1 to 50 kJ/cm² suggests a non-linear *fs*-driven sublimation process in this range. Beyond the 50 kJ/cm² transition, further dose increases, even by an order of magnitude (×10) had limited effect, with *w* and *d* increasing only by ~20 – 30% at 500 kJ/cm². These two regimes are highlighted in Figure 3, in light blue (<50 kJ/cm²) and light red (>50 kJ/cm²) overlays marking the ~50 kJ/cm² transitional point.

## 3.3 Total material volume removed

Figure 4(a) and Figure 4(b) show the variation in total volume removed ($V$, in $\mu m^3$) as a function of laser energy dose (kJ/cm²), plotted on linear and $\log_{10}(x)$ scales, respectively.

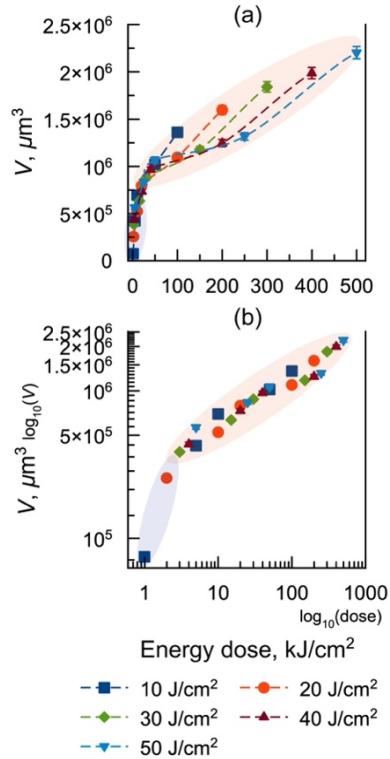

Figure 5: Variation in total volume removed ($V$, in $\mu m^3$) for photoablated tracks in diamond, plotted on (a) linear and, (b) $\log_{10}$ scale as a function of 30-*fs* laser energy dose. Low-dose (<50 kJ/cm$^2$; blue) and high dose (>50 kJ/cm$^2$; orange) ablation regimes are indicated with overlays.

The $V$ followed a trend consistent with observed changes in ablated track's $w$ and $d$ shown in Figure 3, marked by a transitional threshold at ~50 kJ/cm$^2$ dose, where removed $V$ remained at or below ~5×10$^6$ $\mu m^3$. Figure 4(a) aligns with previous findings and reinforces the trend. The 50 kJ/cm$^2$ transition is particularly evident in Figure 5(b) with data plotted on $\log_{10}$ scale which highlights proportional relationships and exponential growth/decay patterns across multiple orders of magnitude. The log-scale representation reveals small but significant changes at lower doses (1 - 2 kJ/cm$^2$), while larger variations become more pronounced above 2 kJ/cm$^2$, especially in datasets with pseudo-exponential DoE parameters (see Section 2.2). As the dose increases from 1 to 50 kJ/cm$^2$, as highlighted in blue, Figure 5(a), the $V$ increases following the same pattern observed for $d$ and $w$ earlier. However, exceeding the 50 kJ/cm$^2$ transition yields only marginal gains in $V$, with only a two-fold increase in $V$ removed despite a ten-fold increase in dose (500 kJ/cm$^2$), making 30-*fs* laser photoablation processes above 50 kJ/cm$^2$ highly inefficient, Figure 5(a). Figure 5(b) further emphasises that significant $V$ removal occurs at or below 2 kJ/cm$^2$, a threshold notably lower than suggested by the linear scale data representation.

These two distinct photoablation regimes arise from the interplay between non-linear ionization phenomena and material response. In the low-dose regime (<50 kJ/cm$^2$, linear scale), multi-photon ionization dominates, efficiently generating $\bar{e}$-hole pairs, leading to rapid $sp^3$ bond-breaking and, efficient material removal. This results in significant $d$ and $w$ increases in irradiated regions. The process remains entirely non-thermal, with energy directly transferred to electrons, driving strong ablation [20, 25]. Notably, non-linear ionisation becomes significant at doses below 2 kJ/cm$^2$, lower than previously observed.

In contrast, the high-dose regime (>50 kJ/cm$^2$) is far less effective for material removal due to plasma shielding from free electron accumulation, as these $\bar{e}$ reflect and absorb incident $fs$-laser energy, modifying local electric fields and reducing ionization efficiency. Additionally, excess energy promotes $sp^2$ amorphization and incomplete $sp^2$-to-$sp^2$ phase conversion, further limiting volume removal efficiency. Beyond 50 kJ/cm$^2$, these effects outweigh the benefits of high energy deposition, marking the transition point where volume removal diminishes despite increased dose.

### 3.4    Material removal rate

The variation in MRR ($\mu$m$^3$/s) as a function of laser energy dose (kJ/cm$^2$) in diamond is shown in in Figure 6(a) linear scale, and Figure 6(b) log$_{10}(x)$ scale. As previously noted, high-precision machining requires an MRR $\geq 0.001$ mm$^3$/s [8], while ultra-high-precision machining is achieved at MRR $\geq 0.0001$ mm$^3$/s, particularity processing advanced carbides and ceramics [9].

The highest overall MRR occurs below the 50 kJ/cm$^2$ threshold, as shown in Figure 6(a). Figure 6(b) further indicates that high-precision standards are met within the 2 - 50 kJ/cm$^2$ range, while ultra-high-precision machining is achieved even at the lowest 1 kJ/cm$^2$ dose. Beyond ~50 kJ/cm$^2$, MRR gains are minimal. This threshold can be reached by either using (i) lower fluences (10 - 30 J/cm$^2$) with higher $n$, or (ii) higher fluences (40 - 50 J/cm$^2$) with lower $n$. Both approaches remain suitable for ultra-high precision machining, provided the total irradiation dose does not exceed ~50 kJ/cm$^2$.

The MRR achieved with 30-$fs$ pulses in this study is comparable to the values reported by Ogawa $et$ $al.$ [26], who used utilised significantly longer 350-$fs$ pulses for machining polycrystalline diamond. This indicates that reducing the $fs$-pulse duration by an order of magnitude does not adversely affect photoablation efficiency. However, MRR declines significantly as irradiation dose increases beyond 100 kJ/cm², Figure 6(b), enforcing the inefficiency of higher energy doses in *fs*-laser machining of diamond.

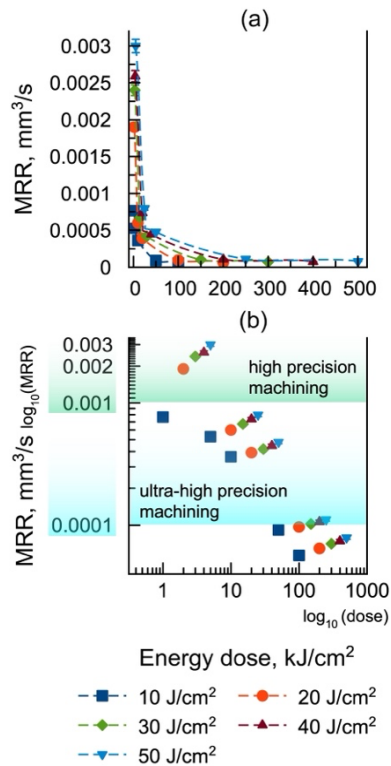

Figure 6: Material removal rate (MRR, µm³/s) as a function of 30-*fs* laser energy dose in diamond, shown on (a) linear scale and (b) log₁₀ scale. High-precision (light green) and ultra-high-precision (turquoise) machining benchmarks are highlighted.

## 3.5    Surface roughness of photoablated tracks

The variation of the average arithmetic surface roughness ($R_a$, *µm*) as a function of 30-*fs* laser energy dose (kJ/cm²) in diamond is shown on a (a) linear, and (b) log₁₀ scales in Figure 7(a) and Figure 7(b), respectively. As previously discussed, in addition to the MRR benchmarks stated in Section 3.3, modern manufacturing precision standards stipulate $R_a$ finish requirements in accordance with ISO 1302 [5] and ASME B46.1 [6]. Specifically, high-precision machining requires an $R_a \leq 0.2$ *µm*, while ultra-high-precision machining demands an $R_a \leq 0.1$ *µm* [7, 27].

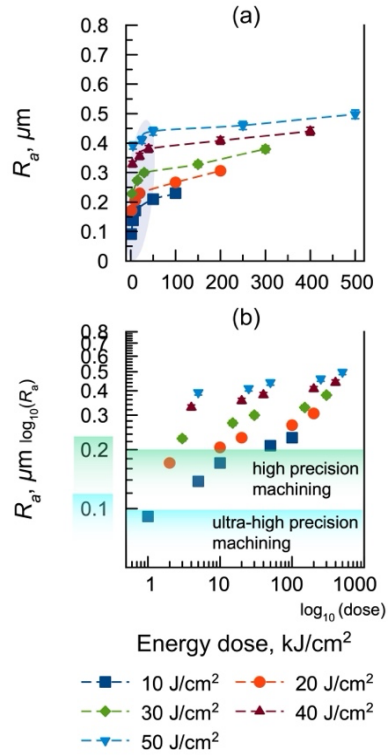

Figure 7: Average arithmetic surface roughness ($R_a$, $\mu m$) as a function of 30-*fs* laser energy dose in diamond, shown on (a) linear scale with the low-dose regime (<50 kJ/cm², blue overlay) and (b) log₁₀ scale, highlighting high-precision (light green) and ultra-high-precision (turquoise) $R_a$ benchmarks.

A lower $R_a$ is generally linked to improved performance, durability, and aesthetic quality, as the minimises stress concentrations, surface defects, and wear. The lowest $R_a$ (0.09 ± 0.005 $\mu m$) was achieved at the lowest irradiation dose (1 kJ/cm²), Figure 7(b), meeting ultra-high-precision machining standards required for Space and Defence applications [28]. $R_a$ <0.2 $\mu m$ was consistently attained at fluences < 10 kJ/cm², satisfying high-precision quality standards [5, 6].

$R_a$ displayed the greatest sensitivity at lower irradiation doses, with sharp decline as the 30-*fs* dose increased from 1 to 10 kJ/cm². The highest $R_a$ (~0.5 $\mu m$) was recorded at 500 kJ/cm², with the most pronounced $R_a$ increase occurring below 50 kJ/cm², Figure 6(a). At doses exceeding 100 kJ/cm², this effect became less significant, denoting that lower fluences (<10 J/cm²) produce the smoothest surfaces.

Notably, our study reports the lowest $R_a$ (sub-0.1 $\mu m$) for laser-machined diamond to date. In comparison, longer *fs*-pulse durations such as 350 *fs* and a 700 *fs* have been shown to produce $R_a$ ~0.2 $\mu$m and 0.4 $\mu$m, respectively, as demonstrated by Ogawa *et al.* [26] and Okuchi *et al.* [29].

Wang *et al.* [30] proposed that the $sp^3$-to-$sp^2$ transition in diamond follows distinct pathways depending on laser pulse duration. With *ns*- and longer pulses, $sp^2$ graphitization propagates vertically into the bulk, forming intercalated $sp^3$-$sp^2$ interfaces post-laser treatment. In contrast, shorter *fs* pulses induce layer-by-layer graphitization confined to the surface, resulting in superior post-ablation surface quality. Wang's atomistic model aligns with our experimental results, where the lowest $R_a$ were achieved using 30-*fs* laser pulses.

### 3.6    Chemical composition of the *fs* micromachined surfaces

The Raman spectra of an *n*-type CVD type-Ib diamond with varying N content (0 to 200 ppm) are shown in Figure 8(a) [31]. A prominent peak at ~1332 cm⁻¹ is observed, along with broad spectral contributions below 1480 cm⁻¹, and near 2100 cm⁻¹, when probed with green light (514 or 532 nm) photons ($\hbar\omega$). The latter two bands are associated with N-related defect features [31-33].

Figure 8(b) presents the Raman spectra of untreated (lime green) and, 30-*fs* laser photoirradiated diamond, comparing the lowest (1 kJ/cm²; dark blue) and highest (500 kJ/cm²; light blue) irradiation doses after the removal of the $sp^2$ graphitised layer. The Figure 8(b) spectra closely resemble those of type-Ib diamond containing 200 ppm N (highlighted in yellow), demonstrating that the intrinsic tetrahedral $sp^3$ phase of diamond remains unaffected by 30-*fs* laser irradiation. This is evidenced by the unchanged core Raman mode at ~1332 cm⁻¹, with its half-width-at-half-maxima (HWHM) remaining well-defined across all laser-treated samples at $\Gamma_{HMHM} = 2.5 \pm 0.2$ cm⁻¹. The 1332 cm⁻¹ mode corresponds to the 1ˢᵗ-order excitation of the triply degenerate $T_{2g}$ optical phonons at the Brillouin zone centre [34], excited by 2.33 or 2.41 eV $\hbar\omega$, consistent with prior findings by Ali *et. al.* [15, 16].

Furthermore, no core Raman *G* or *D* modes are detected, confirming the absence of graphitic fractions within the amorphous carbon matrix [22]. The *G* mode, originating from zone-centre ($\Gamma$-point) phonons of the Brillouin zone, corresponds to the doubly degenerate $E_{2g}$ vibrational mode, where in-plane stretching of $sp^2$-bonded carbon in hexagonal rings or chains dominates [35]. Typically, the *G* peak is resonant between 1530 cm⁻¹ and 1600 cm⁻¹ when excited by 2.33 or 2.41 eV $\hbar\omega$ light. Its absence in the Raman spectra indicates the absence of residual lattice disorder associated with $sp^2$ amorphization or structural defects in the 30-*fs* photoablated regions. Additionally, no broadening is observed around 1350 cm⁻¹, where the Raman-active *D* mode typically appears [36]. The *D* mode, associated with aromatic $sp^2$ ring breathing vibrations, is also absent, further confirming that 30-*fs* laser irradiation preserves the $sp^3$ lattice structure. The complete removal of $sp^2$ fractions through sonication suggests these fractions are soft and aromatic, rather than amorphous, making them easily dislodged during the sonication process.

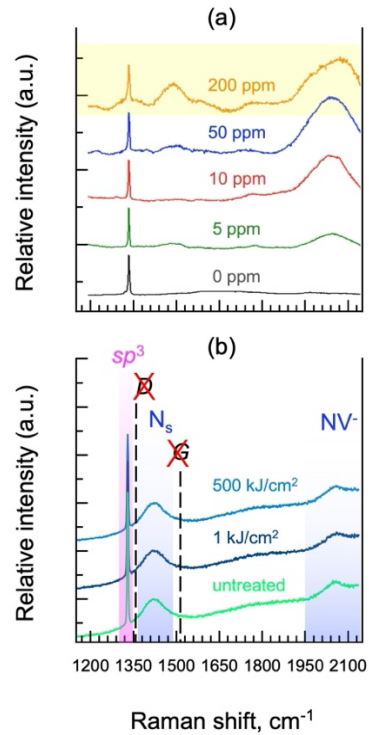

Figure 8. Raman spectra of (a) *n*-type CVD type-Ib diamond with varying N content (200 ppm highlighted in yellow, adapted from [31]), and (b) untreated *vs.* 30-*fs* laser-processed diamond, comparing the lowest (1 kJ/cm$^2$; dark blue) and highest (500 kJ/cm$^2$; light blue) irradiation doses, and their vibrational modes.

In addition to the core $T_{2g}$ peak at 1332 cm$^{-1}$, two additional peaks at ~1420 cm$^{-1}$ and 2060–2100 cm$^{-1}$ in N-doped CVD diamond are attributed to N-related defects and their associated vibrational modes. The 1420 cm$^{-1}$ peak corresponds to localized vibrational modes (LVMs) of N impurities, such as single substitutional nitrogen (N$_s$) or A-centres (paired substitutional N atoms). The presence of this LVM N$_s$ peak indicates the incorporation of substitutional N impurities into the lattice during CVD growth.

The 2060–2100 cm$^{-1}$ peak is linked to the zero-phonon line (ZPL) of N-vacancy (NV) centres or vibrational modes of N-vacancy-hydrogen (NV(H)) complexes [37]. Since hydrogen is widely used in the CVD diamond growth process, it can be incorporated into the lattice, including in

*n*-type CVD diamonds [38]. The presence of the NV(H) peak suggests interactions between NV defects and the $sp^3$ lattice, which is characteristic of CVD-grown crystals with high N content, such as those containing ~200 ppm N.

Both peaks are detectable using optical excitation over a broad wavelengths range (400 - 575 nm, 3.0 – 2.16 eV $\hbar w$), and have been extensively studied by Smith *et al.* [31], Jani *et al.* [33], Zhu *et al.* [39], and others [32, 40]. Notably, the 2100 cm$^{-1}$ peak serves as a relative indicator of N concentration in CVD-grown diamond, as shown in Figure 8(a) [31] and further reported by Smith *et al.* [31].

The fabrication of NV centres typically involves annealing diamond above 700 °C to relieve lattice strain and enhance NV centre mobility [41]. Our results indicate that 30-*fs* laser irradiation does not degrade or adversely affect the $sp^3$ lattice, unlike conventional *ns*- or *ps*- laser irradiation, which is often associated with NV centre annealing [42] and $sp^3$-phase amorphization. This highlights the potential of sub-50-*fs* laser pulses to minimise NV centre thermalisation while preserving the $sp^3$ phase crystallinity and the degree of order in diamond. Additionally, the predominantly non-thermal nature of the 30-*fs* laser process ensures that NV-containing diamond workpieces can be precisely machined into smaller components, without compromising NV site integrity.

## 4    CONCLUSIONS

A 30-*fs*, 800 nm, 1 kHz femtosecond laser was used to photoablate *n*-type CVD diamond with a radiant energy dose range of 1 - 500 kJ/cm², with laser fluences of 10 - 50 J/cm² and pulse counts from 100 to 10,000. The DoE was set up to analyse non-linear trends in net material volume removal, MRR and $R_a$, aiming to identify the optimal irradiation conditions that maximise material removal while maintaining high-precision surface quality irradiating diamond above the photoablation threshold.

Our findings demonstrate that 30-*fs* laser photoablation:

- Achieves high-quality micromachining at $R_a$ <0.1 $\mu$m, meeting both high-precision ($R_a$ <0.2 µm) and ultra-high-precision ($R_a$ <0.1 µm) machining standards with no surface or heat-affected zone (HAZ) damage. At the lowest dose of 1 kJ/cm² (10 J/cm² fluence, 100 pulses), an $R_a$ of 0.09 $\mu$m was obtained, satisfying ultra-high precision machining criteria. Doses below 10 kJ/cm² consistently met high-precision machining requirements across various fluence and pulse combinations.

- Achives maximum photoablation efficiency below 50 kJ/cm², beyond which material removal yields diminish, making irradiation above this threshold highly inefficient.
- Preserves the $sp^3$ phase of diamond, confirmed by the unchanged $T_{2g}$ Raman mode at 1332 cm$^{-1}$ and a consistent $\Gamma_{HMHM}$ of $2.5 \pm 0.2$ cm$^{-1}$ across all laser-treated samples. The absence of Raman $G$ or $D$ modes after 30-$fs$ laser treatment and sonication confirms the absence of residual $sp^2$ inclusions, graphitized fractions, or structural disorder.
- Enables safe machining of NV centre-containing diamond, as the non-thermal nature of the 30-$fs$ laser process ensures NV site preservation.

Additionally, diamond's response to 30-$fs$ laser irradiation is highly non-linear, particularly around the 50 kJ/cm² threshold, where non-linear ionization phenomena, plasma shielding, and material response affect photoablation dynamics at surface. At 50 kJ/cm², these effects begin to counteract the benefits of increased energy deposition, marking a critical efficiency limit.

Overall, these findings highlight the potential of 30-$fs$ laser photoablation as an effective, damage-free, and HAZ-free precision machining solution for diamond-based applications.

## CONFLICT OF INTEREST

The authors have no conflict of interest to disclose.

## CRediT AUTHORSHIP CONTRIBUTION STATEMENT

MR: conceptualisation, methodology, formal analysis, visualisation, writing – original draft, writing - review & editing, data curation, resources, supervision, project administration. BA: investigation, data collection, writing - review & editing. IVL: conceptualisation, methodology, writing - review & editing, resources, supervision, project administration.

All authors gave the final approval for publication and agreed to be held accountable for the work performed herein. The corresponding author attests that all listed authors meet authorship criteria and, that no others meeting the criteria have been omitted.

## ACKNOWLEDGEMENTS

The authors acknowledge the facilities, scientific support, and technical assistance provided by the Australian Microscopy & Microanalysis Research Facility at the Centre for Microscopy and Microanalysis, University of Queensland, and the Central Analytical Research Facility at